\documentclass[11pt,a4paper]{article}
\usepackage[left=15mm, top=10mm, right=15mm, bottom=20mm, nohead=true, nofoot=true]{geometry}

\usepackage[utf8]{inputenc}
\usepackage{authblk}
\usepackage{indentfirst}
\usepackage{misccorr}
\usepackage{graphicx}
\usepackage{amsmath}
\usepackage{amssymb}
\usepackage{multicol}
\usepackage{bm}
\usepackage{longtable}
\usepackage{pdflscape}
\usepackage{multirow}
\usepackage{adjustbox}
\usepackage{amsfonts}
\usepackage{ifthen}
\usepackage{fancyhdr}
\usepackage{setspace}
\usepackage{xcolor}
\usepackage[round,authoryear,comma,sort&compress,numbers]{natbib}
\usepackage[toc,title,page]{appendix}
\usepackage[hidelinks]{hyperref}
\usepackage{url}

\hypersetup{
    colorlinks=true,
    linkcolor=green,
    citecolor=blue,
    filecolor=magenta,
    urlcolor=cyan,
    pdftitle={Sharelatex Example},
    pdfpagemode=FullScreen,
}

\fancypagestyle{plain}{\chead{\texttt{\textcolor{brown}{Communications of BAO, Vol. ?, Issue ?, 20??, pp. ???-???}}}}
\title{\textbf{The influence of environmental effects on Type Ia Supernovae Standardization}}
\author[1]{A.~Yu. Baluta \thanks{nast0307@mail.ru, corresponding author}}
\author[2]{M.~V. Pruzhinskaya\thanks{pruzhinskaya@gmail.com}}
\author[2]{P. Rosnet}
\author[2]{N. Pauna}
\affil[1]{\scriptsize Faculty of Physics, Lomonosov Moscow State University, Moscow, Russia}
\affil[2]{\scriptsize Laboratoire de Physique de Clermont,  Université Clermont Auvergne, IN2P3/CNRS, France}

\def\apj{Astrophys.~J. }

\def\aap{Astron.~Astrophys. }

\def\apjl{Astrophys.~J.~Lett. }
\def\apjs{Astrophys.~J.~Suppl.~Ser. }
\def\aj{Astron.~J. }
\def\mnras{Mon.~Not.~R.~Astron.~Soc. }

\def\sovast{{\textbackslash}sovast}

\begin{document}
\pagestyle{empty}
\newpage
\pagestyle{fancy}
\label{firstpage}
\date{}
\maketitle

\begin{abstract}
Type Ia Supernovae (SNe~Ia) are used as reliable cosmic distance indicators and their standardization is necessary for a more accurate measurement of the cosmological parameters of the Universe. However, the Hubble diagram still shows some intrinsic dispersion, potentially influenced by the supernova's environment. In this study, we reproduce the Hubble diagram fit for the Pantheon supernova sample, and also investigate the possibility of introducing various standardization equations for supernovae exploded in early- and late-type galaxies. We analyze 330 Pantheon SNe~Ia to study how host galaxy morphology affects their standardization. We find that SNe~Ia hosted by early-type galaxies have different standardization parameters compared to those hosted by late-type galaxies. We conclude that correcting supernova luminosity for host galaxy morphology is essential to perform the precise cosmological analysis. 

\end{abstract}
\emph{\textbf{Keywords:} supernovae: general; cosmology: observations, cosmological parameters, distance scale}

\section{Introduction}
Type Ia supernovae (SNe~Ia) are known as standard candles or indicators of cosmological distances. In 2011, Saul Perlmutter, Brian Schmidt and Adam Riess received the Nobel Prize in Physics ``for the discovery of the accelerating expansion of the Universe through observations of distant supernovae''~\citep{1998AJ....116.1009R, 1999ApJ...517..565P}. In fact, it would be more correct to call SNe~Ia standardizable objects. As observational data accumulated,~\cite{1938ApJ....88..285B} discovered the universality of their light curves: they show a rapid increase to the maximum in 2-3 days and then a drop of $\sim3^m$ for every $25-30$ days.

In the 1970s, B.~W. Rust and Yu.~P. Pskovskii independently revealed an empirical relationship linking the supernova luminosity with its light curve parameters~\citep{1974PhDT.........7R,1977SvA....21..675P}. After that, many models of standardization of Type Ia supernovae were proposed using various parameters of the light curve. In our work, we use one of such models called the SALT2 standardization model~\citep{2007A&A...466...11G} with two light-curve parameters: the color, $c = (B-V)_{MAX} - \langle B-V \rangle$, and the stretch, $x_1$, of the supernova's light curve. The observed distance modulus $\mu$ taking into account the standardization equation is presented as follows:
\begin{equation}\label{baluta:eq1}
    \mu = m_{B} - M_{B} + \alpha x_{1} - \beta c,
\end{equation}
where $m_{B}$ is the apparent magnitude at maximum in  $B$-band, $\alpha$, $\beta$, $M_{B}$ are the parameters of the standardization equation. In the recent cosmological analysis ~\citep{2018ApJ...859..101S}, host stellar mass correction ($\Delta_M$) and correction for distance biases due to intrinsic scatter and selection effects ($\Delta_B$), are taken into account. 

These corrections are suitable yet not ideal as they do not
completely get rid of the distance modulus dispersion on the Hubble diagram. There is an intrinsic supernova luminosity dispersion left, which for one of the most recent cosmological analyses is $\sim0.1^m$ \citep{2022ApJ...938..110B}. We believe this spread may be related to environmental effects. 

The goal of this work is to reproduce the Hubble diagram (supernova's distance modulus vs. redshift) for the Pantheon SN sample~\citep{2018ApJ...859..101S}, as well as to improve the standardization equation taking into account the supernova environment. We investigate the possibility of using different standardization equations for two populations of supernovae: those that exploded in galaxies of early (E-S0/a) and late (Sa-Sd, Ir) morphological types.

\section{Data and methodology}

Our research is based on a comprehensive dataset from the Pantheon sample~\citep{2018ApJ...859..101S}. This sample consists of $1048$ SNe~Ia in the range of  $0.01 < z < 2.3$ discovered by various surveys including Pan-STARRS1 (PS1; \citealt{2014ApJ...795...44R, 2014ApJ...795...45S}), Sloan Digital Sky Survey (SDSS; \citealt{2019BAAS...51g.274K}, Supernova Legacy Survey (SNLS; \citealt{2011ApJS..192....1C, 2011ApJ...737..102S}) and Hubble Space Telescope surveys (HST; \citealt{2007ApJ...659...98R, 2012ApJ...746...85S}). To ensure transparency and reproducibility, we sourced the data from GitHub\footnote{\url{https://sites.astro.caltech.edu/palomar/about/telescopes/oschin.html}} and collected coordinates, redshift ($z_{\rm CMB}$, $z_{\rm HD}$), host galaxy mass, stretch and color parameters, distance modulus, and associated errors for each supernova. We adopt the morphological classification of host galaxies for 330  supernovae from~\citealt{2020MNRAS.499.5121P}. All morphological types of hosts are divided into two general groups: early-type galaxies and late-type galaxies. Thus, $91$ SNe Ia from Pantheon sample exploded in early-type galaxies and $240$ SNe have late-type hosts.

It is known from the literature~\citep{2010ApJ...722..566L, 2006ApJ...648..868S, 2010MNRAS.406..782S, 2003MNRAS.340.1057S, 2020MNRAS.499.5121P, 2018A&A...615A..68R, 2020MNRAS.492..848A} that supernova luminosity depends on the host galaxy parameters (for example, metallicity, stellar mass, star formation rate, local color, galactocentric distance, etc.). In our work we study the influence of host galaxy morphology on the cosmological analysis.

\section{Hubble diagram}

\begin{figure}[ht]
    \centering
    \includegraphics[width=0.6\textwidth]{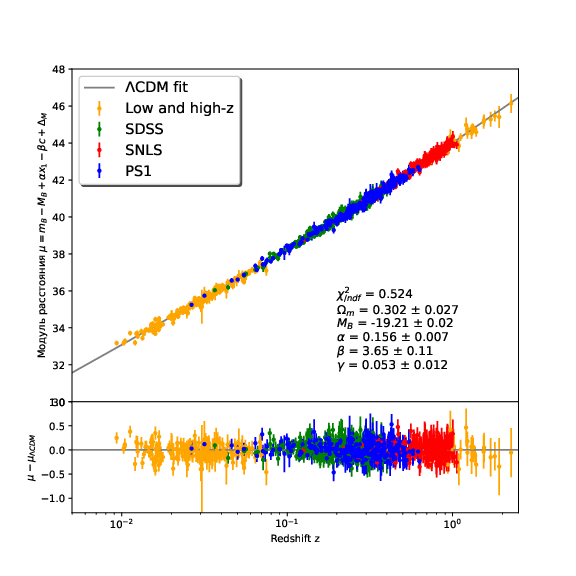}
    \caption{Hubble diagram plotted with all Pantheon supernovae.}
    \label{baluta:fig1}
\end{figure}

To begin with, it is necessary to reproduce the Pantheon Hubble diagram (see Fig.~\ref{baluta:fig1}). Assuming the flat $\Lambda$CDM cosmology, we found standardization parameters of supernovae ($\alpha$, $\beta$, $M_{B}$) and cosmological parameter ($\Omega_{m}$) by minimizing the Hubble residuals. The value of the Hubble constant corresponds to the one recently measured by~\citet{2022ApJ...934L...7R}: $H_{0}$ = 73.04~km~s$^{-1}$~Mpc$^{-1}$.

The results of the fit are presented in the Table~\ref{baluta:tab1} and are in a good agreement with original work of~\citet{2018ApJ...859..101S}. 

\begin{table}
\caption{Comparison of the obtained values of the free parameters of the Hubble fit diagram with the work~\cite{2018ApJ...859..101S} for $H_{0}$ = 73.04~km~s$^{-1}$~Mpc$^{-1}$.}
\begin{center}
\label{baluta:tab1}
\begin{tabular}{|c|c|c|}
\hline
Parameter & This work & \citet{2018ApJ...859..101S} \\
\hline

$\Omega_{M}$ & $0.302 \pm 0.028$ & $0.298 \pm 0.022$\\

$M_{B}$ & $-19.204 \pm 0.016$ & ---  \\

$\alpha$ & $0.155 \pm 0.007$ & $0.157 \pm 0.005$ \\

$\beta$ & $3.687 \pm 0.109$ & $3.689 \pm  0.089$ \\

$\gamma$ & $0.055 \pm 0.012$ & $0.054 \pm 0.009$ \\
\hline
\end{tabular}
\end{center}
\end{table}

\section{Hypothesis of different standardization equations}

~\citet{2020MNRAS.499.5121P} found a dependency between the stretch parameter $x_1$ and the morphological type of the host galaxy of supernova. It was concluded that supernovae in early-type galaxies were characterized by lower values of the stretch parameter compared to supernovae in late-type galaxies. 

It is interesting to see whether the inclination angles of the dependence of $M_B^\star$ (see Eq.~\ref{baluta:eq3}) -- the standardized absolute magnitude -- on the color and stretch parameters for the two supernova populations are different. In case of observing a characteristic fracture on the graph, we can conclude that it is worth to introduce various standardization equations for supernovae in various types of galaxies:
\begin{equation}\label{baluta:eq3}
  M_B^\star = M_B - \alpha_{1} x_{1,PA} - \alpha_{2} x_{1,SF} + \beta_{1} c_{PA} + \beta_{2} c_{SF}\,,
\end{equation}
where $\alpha_{1} x_{1,PA}$ and $\beta_{1} c_{PA}$ are related to early-type galaxies, and $\alpha_{2} x_{1,SF}$ and $\beta_{2} c_{SF}$ to the late-type ones.

\begin{figure}
    \centering
    \includegraphics[width=\textwidth]{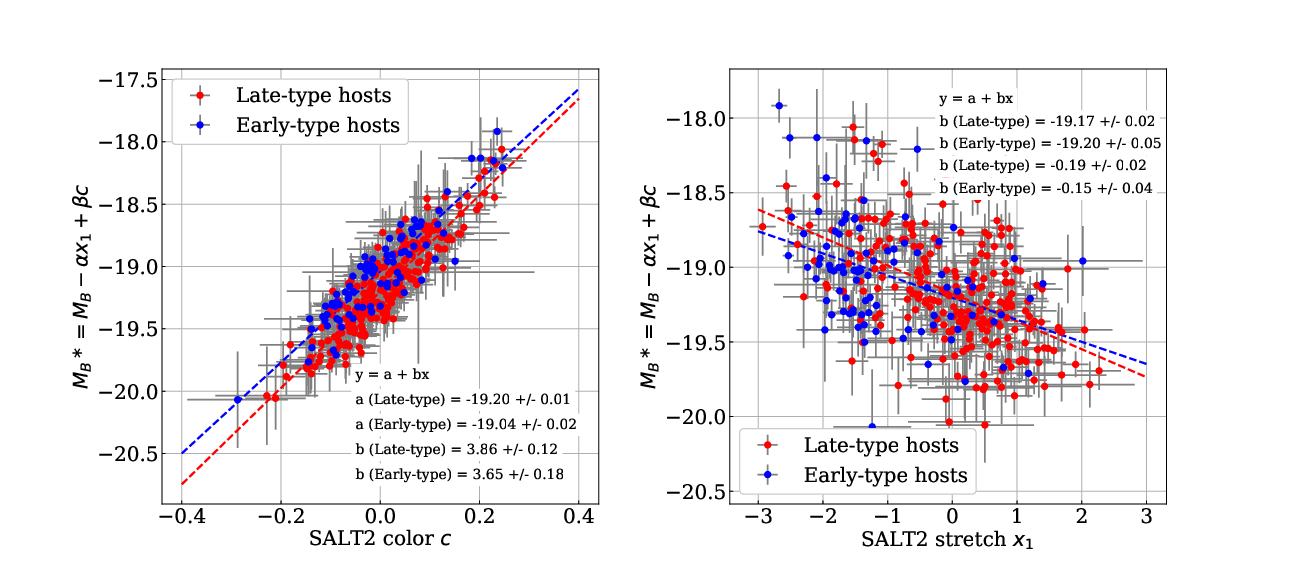}
    \caption{${M_B}^\star$ vs. SALT2 parameters. SNe in late-type galaxies are shown with red colour and blue colour represents SNe in early-type galaxies.}
    \label{baluta:fig2}
\end{figure}

We plotted 330 SNe~Ia with known host morphological types on the Fig.~\ref{baluta:fig2} of the dependence of $M_B^\star$ on the parameters of color and stretch. The $\alpha$ and $\beta$ values were taken from the Table~\ref{baluta:tab1}. The hypothesis is it might be possible to present two different standardization equations for these groups of supernovae and thus we performed linear approximation for them. We do not observe significant differences in the mean values of the slope angle for these two supernova populations. However, on the left subplot we see a shift in magnitudes between two host environments. On the right subplot we do not observe a shift but we can distinguish two clusters, indicating that SNe with smaller $x_1$ are mainly hosted by early-type galaxies.

\section{Supernovae standardization for different groups}

Since we are investigating the features of the supernova environment, we separately approximate the Hubble diagrams for supernovae which exploded in early-type galaxies and for ones exploded in late-type galaxies. We perform a Hubble diagram fit for these two subsamples of supernovae without any environmental correction. The cosmological parameter $\Omega_m$ is fixed to the value given in the Table~\ref{baluta:tab1}. 

\begin{figure}[ht]
    \centering
    \includegraphics[width=3.5in]{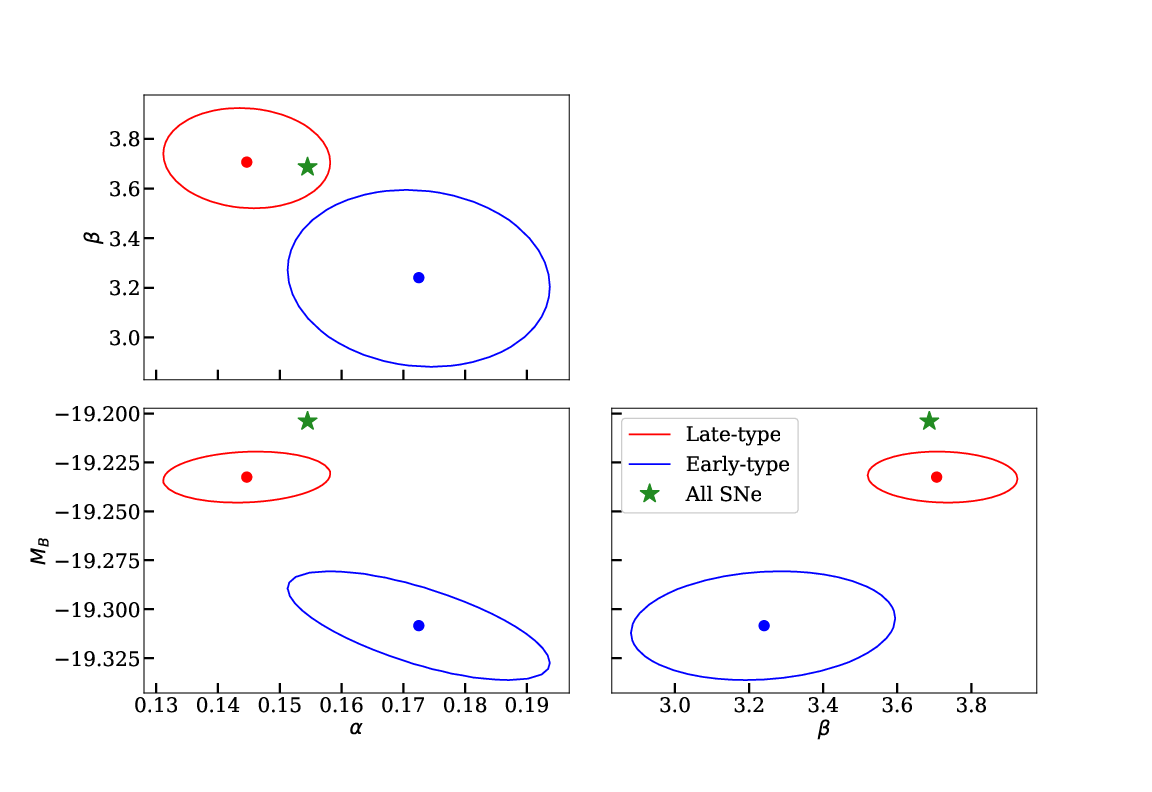}
    \includegraphics[width=3.5in]{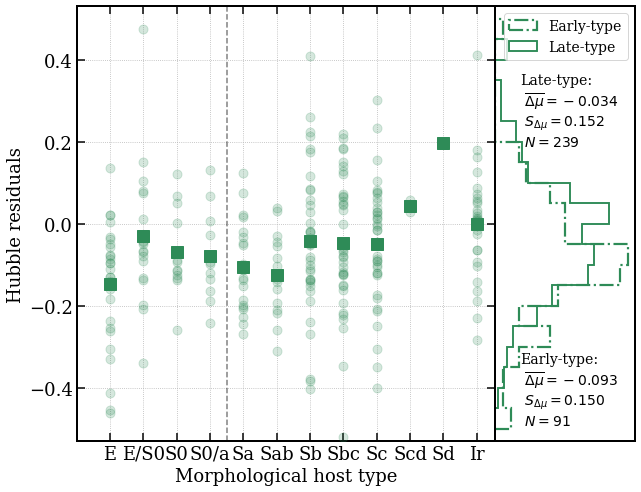}    
    \caption{\emph{Left panel:} Joint confidence contours (1-$\sigma$) in three-parameter plots of $\alpha$, $\beta$, $M_B$ for fits where the supernovae are split according to the host galaxy morphology. Red ellipses represent SNe in star-forming environment and blue ellipses represent SNe in passive environment. Dots are the central values for each subsample and green stars represent values for the whole Pantheon fit. The density of matter is fixed to the value $\Omega_{m} = 0.302$ for the full Pantheon fit ($N_{SNe}=1048$). \\\emph{Right panel:} Hubble residuals for 330 SNe with known host morphology types from Pantheon SNe~Ia sample. The squares denote the mean values $\overline{\Delta\mu}$ in each morphological bin. The right subplot is the normalised histogram of $\Delta\mu$ distribution for early-type and late-type morphological supernovae groups.}
    \label{baluta:fig3}
\end{figure}

The results in the form of joint confidence contour plots containing correlations of the standardization parameters within 1-$\sigma$ are presented on the left panel of Fig.~\ref{baluta:fig3}. The blue color indicates the results of approximation for supernovae in old and passive environment, and red color states for supernovae in young and active environment. Green stars represent standardization parameters obtained for the entire sample of 1048 Pantheon supernovae. At the same time, for two populations, we observe obvious discrepancies in the values of absolute magnitudes $M_B$ depending on color and stretch parameters of the supernovae light curves.

Thus, we state that without any additional environmental correction the standardization procedure is incomplete. This can also be seen on the Hubble residuals plot. On the right panel of Figure~\ref{baluta:fig3} we show the dependence between the Hubble diagram residuals $\Delta\mu = \mu - \mu_{\rm model}$ and host morphology. SNe in early-type galaxies (old environment) possess slightly smaller values of $\overline{\Delta\mu}$ compared to SNe in late-type galaxies (star-forming environment), which indicates that environmental effects need to be taken into account to fulfil the intrinsic dispersion.

\section{Discussion and conclusions}

In the course of our study, we successfully approximated the Hubble diagram for the Pantheon sample of 1048 Type Ia supernovae. This was done both for the entire sample as a whole and for individual groups of supernovae, depending on the type of environment -- star-forming or passive. 

However, the question of introducing various standardization parameters that take into account supernovae environment remains open. The figures showed that the average absolute value of the luminosity of supernovae at the maximum differ depending on the host galaxy type. In particular, supernovae in late-type galaxies have large values of the stretch parameter compared to supernovae in the passive environment. This leads to a shift depending on the color (see Fig.~\ref{baluta:fig2}). In the future, we plan to continue exploring the possibility of introducing various standardization equations for different populations of supernovae: not only the host galaxy morphological type, but also the galactocentric distance can be considered as a criterion for attribution to a particular population.

We confirm that environmental correction should be introduced in order to minimize the luminosity dispersion on the Hubble diagram.

The code to perform the described analysis is available at \texttt{GitHub}\footnote{\url{https://github.com/pruzhinskaya/SN_environment}}. 

\scriptsize
\bibliographystyle{ComBAO}
\nocite{*}
\bibliography{ComBAO}

\end{document}